\documentclass[runningheads]{llncs}
\usepackage[T1]{fontenc}

\usepackage{rotating}
\usepackage{graphicx}
\usepackage{tabularx}
\usepackage{booktabs}
\usepackage{caption}
\usepackage{makecell}
\usepackage{xcolor}

\begin{document}

\title{Interdependent Privacy in Smart Homes: Hunting for Bystanders in Privacy Policies}

\titlerunning{Interdependent Privacy in Smart Homes}

\author{Shuaishuai Liu\and
Gergely Acs
\and
Gergely Bicz\'ok
}

\institute{CrySyS Lab, Budapest Univ. of Technology and Economics\\
\email{\{sliu,acs,biczok\}@crysys.hu}}

\maketitle              

\begin{abstract}
Smart home devices such as video doorbells and security cameras are becoming increasingly common in everyday life. While these devices offer convenience and safety, they also raise new privacy concerns: how these devices affect others, like neighbors, visitors, or people passing by. This issue is generally known as \emph{interdependent privacy}, where one person's actions (or inaction) may impact the privacy of others, and, specifically, \emph{bystander privacy} in the context of smart homes. Given lax data protection regulations in terms of shared physical spaces and amateur joint data controllers, we expect that the privacy policies of smart home products reflect the missing regulatory incentives.
This paper presents a focused privacy policy analysis of 20 video doorbell and smart camera products, concentrating explicitly on the bystander aspect. We show that although some of the vendors acknowledge bystanders, they address it only to the extent of including disclaimers, shifting the ethical responsibility for collecting the data of non-users to the device owner. In addition, we identify and examine real-world cases related to bystander privacy, demonstrating how current deployments can impact non-users. Based on our findings, we analyze vendor privacy policies in light of existing legal frameworks and technical capabilities, and we provide practical recommendations for both policy language and system design to enhance transparency and empower both bystanders and device owners.

\end{abstract}

\section{Introduction}
\label{sec:intro}
Privacy is traditionally understood as an individual right -- the ability to control access to one's personal information~\cite{stone1983field}. 
Application and service providers are obligated by data protection regulations to properly process, store, and share personal data with the consent of the users. However, in today's interconnected world, the data in each user's ``hands'' is not necessarily their own. Technologies such as social media, third-party app platforms, and smart homes have created environments where one person's privacy-related decisions often affect others. This phenomenon is known as \emph{interdependent privacy} (IDP)~\cite{DBLP:conf/fc/BiczokC13}; fundamentally, some individuals (whether users of a given service or non-users) may suffer from the negative externalities created by another individual's decision to share their data. Earlier works in this topic focused mostly on social and cloud apps, location sharing, and genomic privacy~\cite{idp_survey}. Keeping up with technological progress, IDP research in the current decade explored the pervasiveness of IDP across app platforms~\cite{liu2021interdependent}, large language models~\cite{zhan2024beyond}, virtual spaces~\cite{o2023privacy}, and, most prominently, smart homes~\cite{saqib2025bystander}. 

Acknowledged as the physical manifestation of IDP by the privacy research community~\cite{saqib2025bystander,kroger2022personal,alshehri2022exploring,frik2025cares}, \emph{bystander privacy} poses a crucial challenge in the domain of smart spaces. Smart devices are often installed and managed by one person (the device owner), but they can collect data on everyone in their environment. Unlike online platforms where users might opt in or adjust privacy settings, bystanders in physical spaces rarely have such options. Their data is collected passively, sometimes even without their awareness -- this creates a power imbalance. The device owner gains increased control and security, while others, who may not even know they are being recorded, bear the privacy costs. In a household using smart assistants like Amazon Echo or Google Nest, conversations among guests or family members may be recorded even if those individuals never interacted with the device\footnote{Note that in such cases Art. 2(2)(c) GDPR (household exemption) may apply.}. The same applies to outdoor smart surveillance systems, which might capture neighboring properties or public sidewalks. 
The blurring of the perceived ``boundary'' between users and bystanders/non-users suggests the need for a new approach for privacy protection, one that considers group dynamics, shared spaces, and overlapping spheres of privacy. Instead of focusing solely on individual consent and control, countermeasures should also account for collective risks and externalities.

To make matters even more complicated, existing legal frameworks, such as the GDPR and CCPA, primarily protect the data subject, while they are not explicit about responsibilities in shared physical environments, particularly when it is unclear who controls the data. Lacking proper regulatory incentives, bystander privacy (and IDP in general) is often overlooked in both system design and compliance efforts.  This lack of explicit regulatory guidance leaves privacy protection to the goodwill of device vendors and service providers, and the reasonably poor awareness of device owners.  As a result, privacy policies become points for understanding how/if bystander privacy concerns are addressed in practice.
A large body of work has emphasized the importance of systematically analyzing privacy policies to assess the extent to which they support transparency and ethical data practices~\cite{VANDERSCHYFF2024104065}. Best practices, challenges, and opportunities for privacy scholars in extracting meaningful insights from privacy policies have also been identified~\cite{DBLP:journals/popets/MhaidliFDHSMWS23}. Prior negative findings of IDP research regarding third-party app platforms~\cite{SYMEONIDIS2018179,DBLP:journals/compsec/LiuB25} and digital address books~\cite{SalehzadehNiksiratUSENIXDAB-IDP} leave us expecting that shortcomings are also evident in the privacy policies of smart device manufacturers. In fact, Alshehri et al. reported briefly that only 2 of 17 device manufacturers mentioned bystander privacy in their privacy policies~\cite{alshehri2022exploring}. Although a recent survey~\cite{saqib2025bystander} summarized the concerns and proposed solutions regarding the impact of smart home technologies on bystander privacy, a deeper analysis of privacy policies of contemporary device manufacturers is yet to be presented.

In this paper, we provide a focused privacy policy analysis of 20 video doorbell and smart camera manufacturers, concentrating specifically on the bystander aspect. Our contributions are twofold. First, we show that although some of the vendors acknowledge bystanders, they seem to consider the issue sufficiently handled by including disclaimers, shifting the responsibility for collecting non-users' data to the device owner. Second, we provide actionable recommendations concerning both the content of privacy policies and technical system design to promote transparency and empower bystanders (and users).

The remainder of this paper is structured as follows. Section~\ref{sec:smarthomecase} reviews the development of smart space products and highlights major incidents and observations related to bystander privacy risks. Section~\ref{sec:policyanalysis} presents our privacy policy analysis framework and key findings regarding how leading smart home vendors address bystander privacy risks. Section~\ref{sec:ppanalysis} details the findings and compiles clear recommendations for useful privacy policies in smart spaces. Finally, Section~\ref{sec: conclusion} concludes the paper.

\section{Smart home devices}
\label{sec:smarthomecase}

Smart home devices are now part of everyday life. Products like video doorbells, security cameras, smart speakers, and motion sensors make homes more comfortable, convenient, and secure. But these devices do not just impact their owners. Anyone who visits, lives nearby, or simply passes a house equipped with these technologies can be recorded or tracked -- often without being aware.

Many devices record continuously, while others are triggered by motion or sound. These features are designed for physical security, but bring up hard questions: Is it good practice for homeowners to record everyone near their property? What happens to those recordings? Who can access them? Most privacy laws and company policies focus on users, not on people who are incidentally recorded.

Privacy policies rarely explain what happens to the data of non-users. They are often complex and written for owners, not for people who have no choice in the matter. As a result, bystander privacy -- the idea that one person's technology choices can compromise another's privacy -- is mostly ignored in both product design and legal frameworks.
In short, smart home technology creates shared privacy risks, but clear boundaries and protections for bystanders are missing.

\subsection{Case law}
\label{sec:caselaw}

To see how these issues play out in the real world, we reviewed a range of real-world incidents and selected three notable cases. Each one demonstrates how common smart home devices can affect the privacy of people who are not users and how regulations are only beginning to address these challenges.

\subsubsection*{Fairhurst v. Woodard (UK)\footnote{\url{https://www.judiciary.uk/wp-content/uploads/2022/07/Fairhurst-v-Woodard-Judgment-1.pdf}}.}

In 2021, a UK court sided with Dr. Mary Fairhurst, who complained that her neighbor’s Ring cameras captured images and audio of her property without her permission. The court agreed, calling it an “unjustifiable invasion of privacy.” The judge noted that, while the cameras were installed for security, their audio and video coverage was excessive -- audio could be recorded from up to 20 meters away. This case set an important precedent for bystander rights in the context of smart surveillance.

\subsubsection*{FTC v. Ring (US)\footnote{\protect\url{https://www.ftc.gov/legal-library/browse/cases-proceedings/2023113-ring-llc}}.}

In 2023, the US Federal Trade Commission (FTC) settled with Ring (an Amazon company) after finding that Ring employees and contractors could access user videos without consent. Some employees even downloaded and viewed customer videos for their own personal use. While the primary issue was internal abuse, the case also showed that bystanders -- neighbors, visitors, delivery workers -- were often recorded and had no control over their data, despite not being Ring customers.

\subsubsection*{TRENDnet IP camera breach (US)\footnote{\protect\url{https://en.wikipedia.org/wiki/In\_re\_TRENDnet,\_Inc.}}.}

In 2012, a major flaw in TRENDnet IP cameras let anyone view live video feeds without a password. Hundreds of private video streams -- from bedrooms to nurseries -- were exposed online. The FTC charged TRENDnet with failing to secure its products, but the breach’s true impact was felt by people who were recorded unknowingly and had no way to protect their privacy or even know they were at risk.

\subsection{Observations across cases}

\subsubsection*{Practical challenges in defining boundaries in smart spaces. }Fairhurst v. Woodard exposes the difficulty of addressing bystander privacy when smart home devices monitor spaces that extend beyond the owner's private areas. While technology enables continuous surveillance, the law struggles to define acceptable boundaries. For example, in dense urban environments such as apartment complexes, it is unclear how video doorbells or cameras should be positioned to respect both security needs and neighbors' privacy. This case highlights the absence of practical guidelines for defining reasonable surveillance practices in different residential contexts.

\subsubsection*{Invisible victims and the limits of legal remedies. }The second and third cases demonstrate the widespread impact of IDP violations: bystanders are routinely recorded, yet remain unaware and uncompensated. Even when privacy breaches are acknowledged by courts or regulators, the direct beneficiaries of legal remedies are almost always the device users -- not the bystanders whose rights were violated. This illustrates both the pervasiveness of IDP problems and the difficulty in ensuring justice for those outside the user-provider relationship.

\subsubsection*{The role and responsibility of device manufacturers. }Across all cases, manufacturers of smart devices are rarely held accountable for the consequences of bystander privacy violations. Companies typically place the burden on owners to comply with laws and inform others, without offering practical support or safeguards for non-users. This raises a critical question: What responsibilities should manufacturers have in mitigating IDP risks? Effective protection will require device providers to move beyond disclaimers (e.g., transferring accountability to device owners) and take a more constructive role in designing products and policies that account for bystander privacy.

\subsection{Illustrative bystander privacy threat model for smart homes}
\label{sec:threat-model}

\begin{figure}[htb]
  \centering
  \includegraphics[width=\linewidth]{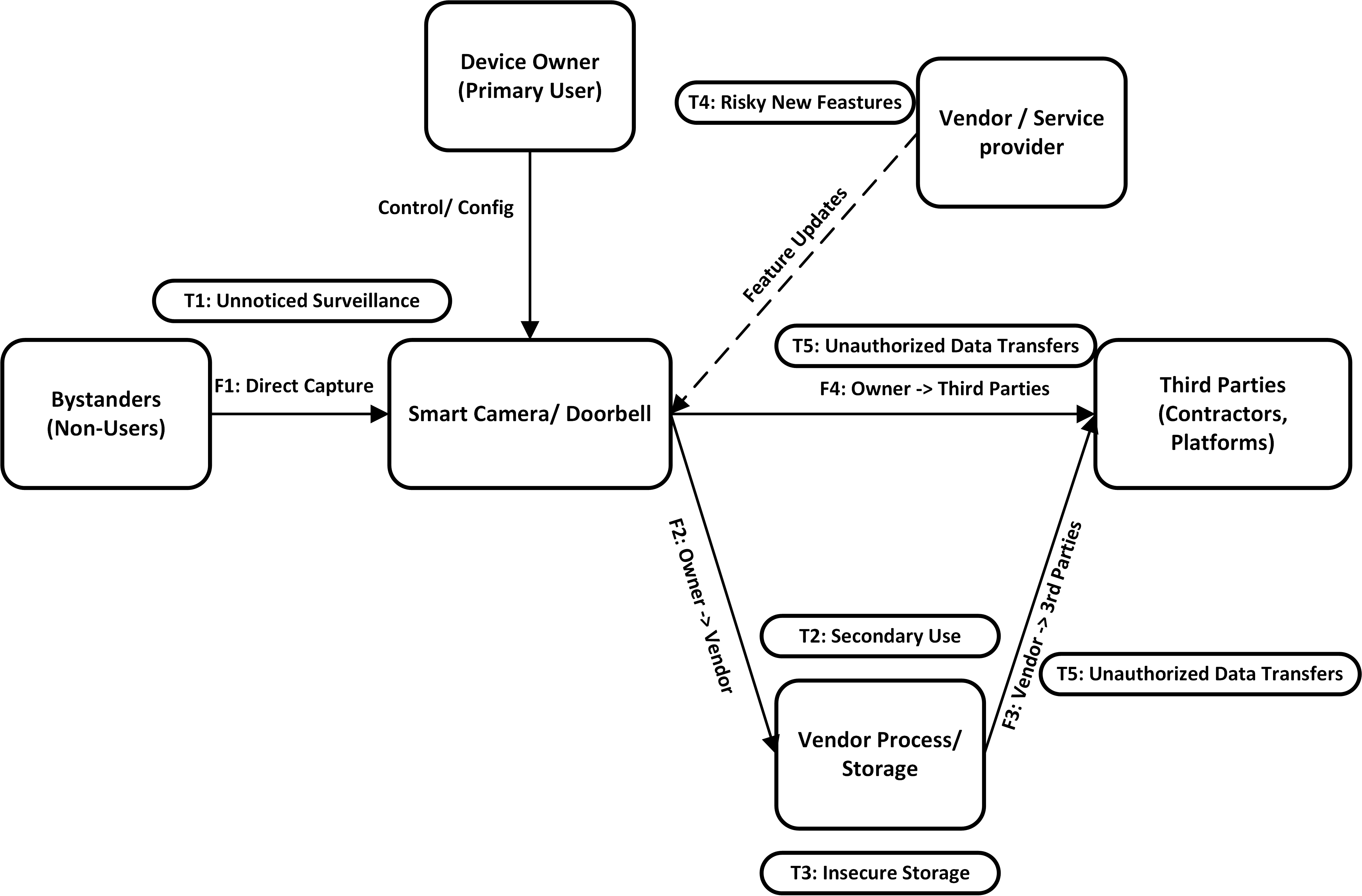}
  \caption{Simplified high-level threat model showing entities, data flows, and threats relevant to bystander privacy.}
  \label{fig:idp-threat-model}
\end{figure}

The case observations in Section~\ref{sec:caselaw} show that IDP in smart homes involves multiple actors, complex data flows, and varied risks. To examine these risks, we define a \emph{simplified, high-level} threat model for bystander privacy in smart home environments. This model is \emph{illustrative rather than exhaustive} and is intended to help readers understand the main entities, data flows, and threat categories discussed in this study (for a comprehensive threat modeling framework see~\cite{liu2025modelinginterdependentprivacythreats}). Fig.~\ref{fig:idp-threat-model} illustrates the model.

\paragraph{Entities.}
\begin{description}
  \item[\textbf{Device Owners (Primary Users).}] Individuals who install and control the smart home device. Some brands, e.g., Amazon Ring, established that the device owner is the data controller and the vendor acts only as data processor\footnote{\url{https://ring.com/eu/en/legal}, see RING GDPR DATA PROCESSING ADDENDUM}.   
  They may also become \emph{de facto} joint controllers\footnote{\url{https://gdpr-info.eu/art-26-gdpr/}} of others’ data~\cite{helberger2010little}.
  \item[\textbf{Bystanders (Non-Users).}] Individuals within the device’s capture range whose personal data is recorded without a contractual relationship to the vendor.
  \item[\textbf{Vendors/Service Providers.}] Companies that manufacture devices, provide cloud services, and set privacy policies.
\end{description}

\paragraph{Data flows.}
\begin{description}
  \item[\textbf{F1: Direct Capture}] Recording of bystanders’ data by device sensors (video, audio, biometrics).
  \item[\textbf{F2: Owner-to-Vendor Transmission}] Upload from device to vendor-managed storage or analytics.
  \item[\textbf{F3: Vendor-to-Third Party Sharing}] Disclosure to contractors, analytics services, or law enforcement.
  \item[\textbf{F4: Owner-to-Third Party Disclosure}] Manual sharing by owners, e.g., on social media.
\end{description}

\paragraph{Threat categories.}
\begin{description}
  \item[\textbf{T1: Unnoticed Surveillance}. ] No effective notice to bystanders about recording.
  \item[\textbf{T2: Secondary Use without Consent. }] Data used beyond the original collection context without bystander approval.
  \item[\textbf{T3: Insecure Storage and Transmission. }] Weak encryption or access control exposes data to unauthorized parties.
  \item[\textbf{T4: Risky Data Practices in New Features. }] New or updated features process or share data in ways that create additional privacy risks, without clear notice, controls, or renewed consent (e.g., enabling continuous recording instead of motion-triggered capture, adding facial recognition or biometric tagging, activating cloud-based person or object detection, integrating with third-party law enforcement portals, or extending data retention by default).
  \item[\textbf{T5: Unauthorized Data Transfers. }] Data sent to other parties without the bystander’s knowledge or permission, including transfers to cloud services, contractors, or integrated platforms.
\end{description}

\paragraph{Roles and responsibilities. }
In this context, the service provider/vendor is either only a data processor or, at a minimum, not the sole controller. Device owners act as intermediaries with partial control but with unclear formal compliance requirements (as amateur controllers) or support. Vendors often shift legal (by proclaiming themselves data processors\footnote{https://ring.com/eu/en/legal}) and ethical responsibilities to owners through disclaimers (see Section~\ref{sec:ppanalysis} for details).

\paragraph{Example.}
A video doorbell captures video of pedestrians (F1), stores the footage in a vendor’s cloud (F2), shares flagged clips with a subcontractor (F3), and a video clip is posted by the owner to a community group (F4). Each stage can trigger different threats (T1--T5).

\section{Privacy policies of smart home devices}
\label{sec:policyanalysis}

Building on the simplified, high-level threat model in Section~\ref{sec:threat-model}, we assess how current smart home privacy policies address the identified entities, data flows, and threat categories (T1--T5). The model serves as an analytical framework, allowing us to map specific policy provisions to concrete risks faced by bystanders. Here, we apply the threat model to evaluate the scope, clarity, and adequacy of countermeasures offered in vendor policies. To fully understand how contemporary smart home companies address privacy risks, it is essential to look beyond technical design and consider the policies that govern how data is collected, processed, and shared. Ideally, privacy policies are not merely legal documents; they could be the primary channel through which companies can articulate their responsibilities, user rights, and risk management strategies\footnote{not explicitly required by the GDPR}~\cite{harkous2018polisis}. Analyzing these policies provides a valuable window into how companies interpret privacy threats, communicate obligations, and (perhaps most importantly) where their policies fall short, especially in the context of emerging issues like bystander privacy.

\begin{table}[t]
\centering
\caption{Mandatory and recommended components of a smart home camera privacy policy with legal bases.}
\label{tab:legal-basis}
\small
\begin{tabularx}{\linewidth}{p{0.35\linewidth} p{0.5\linewidth} p{0.15\linewidth}}
\toprule
\textbf{Policy} & \textbf{Legal Basis} & \textbf{Req.} \\
\midrule
Controller identity \& contact & Art. 13(1)(a), 14(1)(a) GDPR & Yes \\
DPO contact & Art. 13(1)(b), 14(1)(b) GDPR & Yes (appl.) \\
Purposes \& lawful basis & Art. 6, 9, 13(1)(c), 14(1)(c) GDPR & Yes \\
Categories of data & Art. 14(1)(d) GDPR; CCPA §1798.110 & Yes \\
Data sources (indirect) & Art. 14(2)(f) GDPR & Yes (appl.) \\
Recipients / 3rd parties & Art. 13(1)(e), 14(1)(e) GDPR; CCPA §1798.115 & Yes \\
International transfers & Art. 13(1)(f), 14(1)(f), 44--49 GDPR & Yes \\
Retention period & Art. 13(2)(a), 14(2)(a) GDPR; CPRA & Yes \\
Data subject rights & Art. 13(2)(b)--(c), 14(2)(c)--(d), 15--21 GDPR & Yes \\
Complaint mechanism & Art. 13(2)(d), 14(2)(e), 77 GDPR & Yes \\
Automated decision-making & Art. 13(2)(f), 14(2)(g), 22 GDPR & Yes (appl.) \\
Children’s data & Art. 8 GDPR; CCPA/CPRA; COPPA (US) & Yes (appl.) \\
Policy updates & Art. 12 GDPR; CCPA §1798.130(a)(5) & Yes \\
\midrule
Bystander privacy & EDPB Video Guidelines & Rec. \\
Role clarification & Art. 26 GDPR & Rec. \\
Bystander rights path & Not mandated & Rec. \\
Privacy by design/defaults & Art. 25 GDPR & Rec. \\
DPIA disclosure & Art. 35 GDPR & Rec. (high risk)\\
\bottomrule
\end{tabularx}
\vspace{-3mm}
\end{table}

\subsection{Policy collection and pre-processing}
We employed an automated collection method that systematically retrieved privacy policies, notices, and supplemental documents\footnote{Companies may include additional clauses or updates in obscure locations, such as Eufy’s community privacy page \url{https://community.eufy.com/privacy} or Wyze’s Supplemental Terms of Service for Friendly Faces \url{https://www.wyze.com/pages/terms-of-service-friendly-faces}. These supplemental documents are not always prominently displayed, which increases the risk that users will overlook important changes or conditions related to data collection and usage.} from official vendor websites across regions. The resulting corpus was substantially larger than the sample reported in our tables. As a lightweight pre-processing aid, we applied keyword-based filtering to highlight passages related to bystander data, interdependent privacy situations, and data subject rights. Crucially, automation was used only for collection and triage: all findings in this paper are based on subsequent manual examination and detailed analysis by the authors to ensure accuracy and context-relevant interpretation.

\subsection{Sampling and reproducibility }
From the dataset compiled ($\approx$100 brands), we selected 20 brands for detailed cross-brand analysis. Selection was guided by four criteria: (i) market presence and share as reported in industry surveys,~\footnote{https://www.grandviewresearch.com/industry-analysis/smart-home-security-camera-market} (ii) product diversity, including both doorbells and indoor/outdoor smart cameras, (iii) regional coverage, with vendors active in both US and EU markets, and (iv) accessibility and stability of privacy policies, including explicit effective dates and archived versions. These criteria ensure that the analyzed vendors are representative of the leading and widely deployed smart home surveillance products.

All results reported in this paper are grounded in the thorough, manual examination of policies, notices, and supplemental documents from the selected vendors. To support transparency and reproducibility, we share our retrieval method, source code, and the complete annotated analysis of the 20 representative brands\footnote{\url{https://cloud.crysys.hu/s/FPS2025}}.

\subsection{Analysis dimensions}
\label{sec:analysis-dimensions}

To ensure a meaningful comparison, our analysis adopts six carefully selected dimensions. These dimensions first capture whether privacy policies adequately protect direct users, and subsequently explore how interdependent privacy, i.e., the privacy of non-users and bystanders, is addressed. 
The dimensions are as follows:

\begin{enumerate}
\item \textbf{Acknowledgment of bystander data}: Does the policy explicitly acknowledge that data collected may include personal information of bystanders and non-users?
\item \textbf{Scope of data collection}: Does the privacy policy specify the types of data collected (e.g., audio, video, biometrics, location, sensor data) and the conditions under which the collection occurs (continuous vs. event-triggered recording, user-initiated actions such as setup or diagnostics, background analytics, etc.).
\item \textbf{User control}: What level of control do device owners have over their data (access, modification, deletion)?
\item \textbf{Data storage and sharing}: Which commitments are stated regarding data security, encryption, retention duration, and third-party sharing?
\item \textbf{Legal compliance}: Does the policy mention compliance with major privacy regulations (GDPR, CCPA, etc.)?
\item \textbf{Bystander notification mechanisms}: Does the policy provide clear, practical mechanisms to inform non-users about data collection, and offer avenues to object or seek redress?
\end{enumerate}
These six dimensions are grounded in both prior literature on interdependent privacy and specific regulatory requirements. Table~\ref{tab:legal-basis} summarizes the legal bases and recommended provisions under GDPR, CCPA, and related guidelines that informed our framework. For example, lawful basis for collection (GDPR Art.~6), retention limits (GDPR Art.~5(1)(e)), and data subject rights (Arts.~15–21) map directly onto our dimensions, while bystander notification mechanisms reflect the recommendations of the European Data Protection Board (EDPB) and national data protection authorities. These dimensions systematically capture both user-focused privacy protections and how policies address bystander privacy, reflecting our dual objectives: first, to verify fundamental protections for direct users, and second, to assess explicit policy considerations for bystanders. Google Nest's privacy policy (see Table~\ref{tab:nest_analysis}) serves as an illustrative example
\footnote{analysis results for the other brands can be found in the supplement at \url{https://cloud.crysys.hu/s/FPS2025}}.

\begin{table}[!t]
\centering
\caption{Google Nest: analysis of bystander privacy dimensions. Source: \url{https://nest.com/legal/privacy-statement-for-nest-products-and-services/}}
\scriptsize
\renewcommand\arraystretch{1.1} 

\begin{tabularx}{\linewidth}{@{}p{1.7cm}
>{\centering\arraybackslash}p{1.2cm}
@{\hspace{0.4cm}}X
@{\hspace{0.4cm}}X@{}}
\toprule
\multicolumn{1}{l}{\textbf{Dimension}} &
\multicolumn{1}{c}{\textbf{Status}} &
\multicolumn{1}{c}{\textbf{Policy Evidence}} &
\multicolumn{1}{c}{\textbf{Analytical Commentary}} \\
\midrule

Acknowledge Bystander Data &
Yes &
``You (not Nest) are responsible for ensuring that you comply with any applicable laws... you may need to display a notice that alerts visitors to your home and/or obtain their explicit consent depending on the purpose and the means of your processing activities.'' &
Nest explicitly acknowledges that users may capture bystander data and emphasizes user responsibility to notify or obtain consent from non-users, indicating awareness of bystander privacy issues. \\ \midrule

Scope of Data Collection &
Yes &
``When you enable the recording or streaming features... we may record and process video and/or audio recordings... environmental sensor data, motion, temperature, ambient light, and facial recognition data.'' &
Nest collects extensive data types including video, audio, facial recognition, environmental data, and device telemetry, illustrating a broad data collection scope. \\ \midrule

User Control &
Yes &
``You are the controller of these data and Nest is a processor... You may delete your recordings... reset the device to defaults... access, amend or delete your personal information.'' &
Users can manage recordings, adjust sharing settings, and remove data, with Nest positioning users as data controllers in key contexts, such as video and facial recognition. \\ \midrule

Data Storage \& Sharing &
Yes &
``We use best-in-class data security tools... Personal information... is encrypted while it is being transmitted... We do not rent or sell our customer lists.'' &
Nest implements encryption, retains user control over retention and deletion, and prohibits personal data sharing for commercial gain. \\ \midrule

Legal Compliance &
Yes &
``Nest Labs Inc. complies with the EU–U.S. Data Privacy Framework... subject to the investigatory and enforcement powers of the U.S. Federal Trade Commission (FTC).'' &
The policy confirms adherence to GDPR, DPF, and CCPA where applicable, and includes formal oversight by regulatory authorities. \\ \midrule

Bystander Notification Mechanisms &
Partial &
``You may need to display a notice... obtain explicit consent... recording and sharing clips that involve other people... may affect individuals’ privacy and data protection rights.'' &
While Nest instructs users to provide notice or gain consent, it lacks embedded mechanisms (e.g., audible alerts, visual signals) to ensure bystanders are informed in practice. \\

\bottomrule
\end{tabularx}
\label{tab:nest_analysis}
%\vspace{-5mm}
\end{table}

We apply a three-tiered classification: \textit{Yes}, \textit{Partial}, and \textit{No}. A designation of \textit{Yes} indicates that the policy fully satisfies the dimension’s criteria, clearly addressing the issue in explicit terms. \textit{No} indicates the complete absence of relevant language or acknowledgment in the policy. \textit{Partial} is used when the policy exhibits some awareness or partial coverage of the issue but fails to comprehensively or directly address it. For example, under the dimension \textit{Acknowledgment of bystander data}, we assess whether the policy explicitly recognizes that data collected by the product may include information about non-users or bystanders. A policy that clearly mentions responsibilities toward bystanders or the implications of capturing non-user data qualifies for a \textit{Yes}. However, if the only related content refers to niche populations (e.g., children) without acknowledging the broader category of bystanders, we assign \textit{Partial}. For instance, a policy stating: \textit{``If we become aware that Personal Information that has been submitted to us relates to a child without the consent of a parent or guardian, we will use reasonable efforts to delete that Personal Information as soon as possible''} indicates that while the policy commits to safeguarding the data of certain vulnerable populations, such as children, it does not generalize these protections to encompass all non-users who may be affected. Another example of a \textit{Partial} label occurs when a policy contains vague or indirect references to third parties without clarifying their data subject status. One such policy notes: \textit{``We will only publicly disclose your personal information in the following cases: (1) With your express consent; ... (6) For the purpose of safeguarding your or any other individual’s life, property and other significant legitimate rights and interests while difficult to obtain your consent.''} Although the mention of \textit{``other individual''} implies some recognition of non-user data, it falls short of explicitly stating that bystander data may be collected and detailing how it is handled. 
Accordingly, this instance should not be classified as \textit{Yes}.

\subsection{Key findings}
\label{sec:policy-analysis}

Applying our analytical framework, we thoroughly reviewed publicly available privacy policies from leading smart doorbell and camera brands. Our key findings are summarized in Table~\ref{tab:privacy_comparison_llncs_final}, which compares each brand across the six dimensions defined above.

\begin{sidewaystable}[!p]
\centering
\caption{Handling bystander privacy: evidence from 20 leading smart doorbell/camera brands}
\small
\renewcommand\arraystretch{1.2}
\begin{tabularx}{\textwidth}{@{}l
>{\centering\arraybackslash}m{2.5cm}
>{\centering\arraybackslash}m{2.3cm}
>{\centering\arraybackslash}m{1.8cm}
>{\centering\arraybackslash}m{2.7cm}
>{\centering\arraybackslash}m{2.3cm}
>{\centering\arraybackslash}m{2.7cm}@{}}
\toprule
\textbf{Brand} &
\makecell{\textbf{Acknowledge} \\ \textbf{Bystander}} &
\makecell{\textbf{Data} \\ \textbf{Collection}} &
\makecell{\textbf{User} \\ \textbf{Control}} &
\makecell{\textbf{Data Storage} \\ \textbf{\& Sharing}} &
\makecell{\textbf{Legal} \\ \textbf{Compliance}} &
\makecell{\textbf{Bystander} \\ \textbf{Protection}} \\
\midrule
Ring (Amazon)       & Yes     & Yes & Yes & Yes & Yes & Partial \\
Google Nest         & Yes     & Yes & Yes & Yes & Yes & Partial \\
Arlo                & Partial & Yes & Yes & Yes & Yes & Partial \\
Eufy (Anker)        & Yes     & Yes & Yes & Yes & Yes & Partial \\
Wyze                & Yes     & Yes & Yes & Yes & Yes & No \\
TP-Link (Kasa)      & Yes     & Yes & Yes & Yes & Yes & No \\
Hikvision           & No      & Yes & Yes & Yes & Yes & No \\
Dahua Technology    & No      & Yes & Yes & Yes & Yes & No \\
Reolink             & Partial & Yes & Yes & Yes & Yes & No \\
EZVIZ               & Partial & Yes & Yes & Yes & Yes & No \\
ASSA ABLOY          & No      & Yes & Yes & Yes & Yes & No \\
Vivint Smart Home   & Yes     & Yes & Yes & Yes & Yes & No \\
SimpliSafe          & Yes     & Yes & Yes & Yes & Yes & Partial \\
Blink (Amazon)      & No      & Yes & Yes & Yes & Yes & No \\
YI Technology       & Yes     & Yes & Yes & Yes & Yes & No \\
Panasonic           & Partial & Yes & Yes & Yes & Yes & No \\
Netatmo             & Yes     & Yes & Yes & Yes & Yes & Partial \\
Bosch Smart Home    & Yes     & Yes & Yes & Yes & Yes & Partial \\
Samsung SmartThings & Yes     & Yes & Yes & Yes & Yes & Partial \\
ADT                 & Partial & Yes & Yes & Yes & Yes & No \\
\bottomrule
\end{tabularx}
\label{tab:privacy_comparison_llncs_final}
\end{sidewaystable}

\begin{figure}[!t]
  \centering
\includegraphics[width=\linewidth]{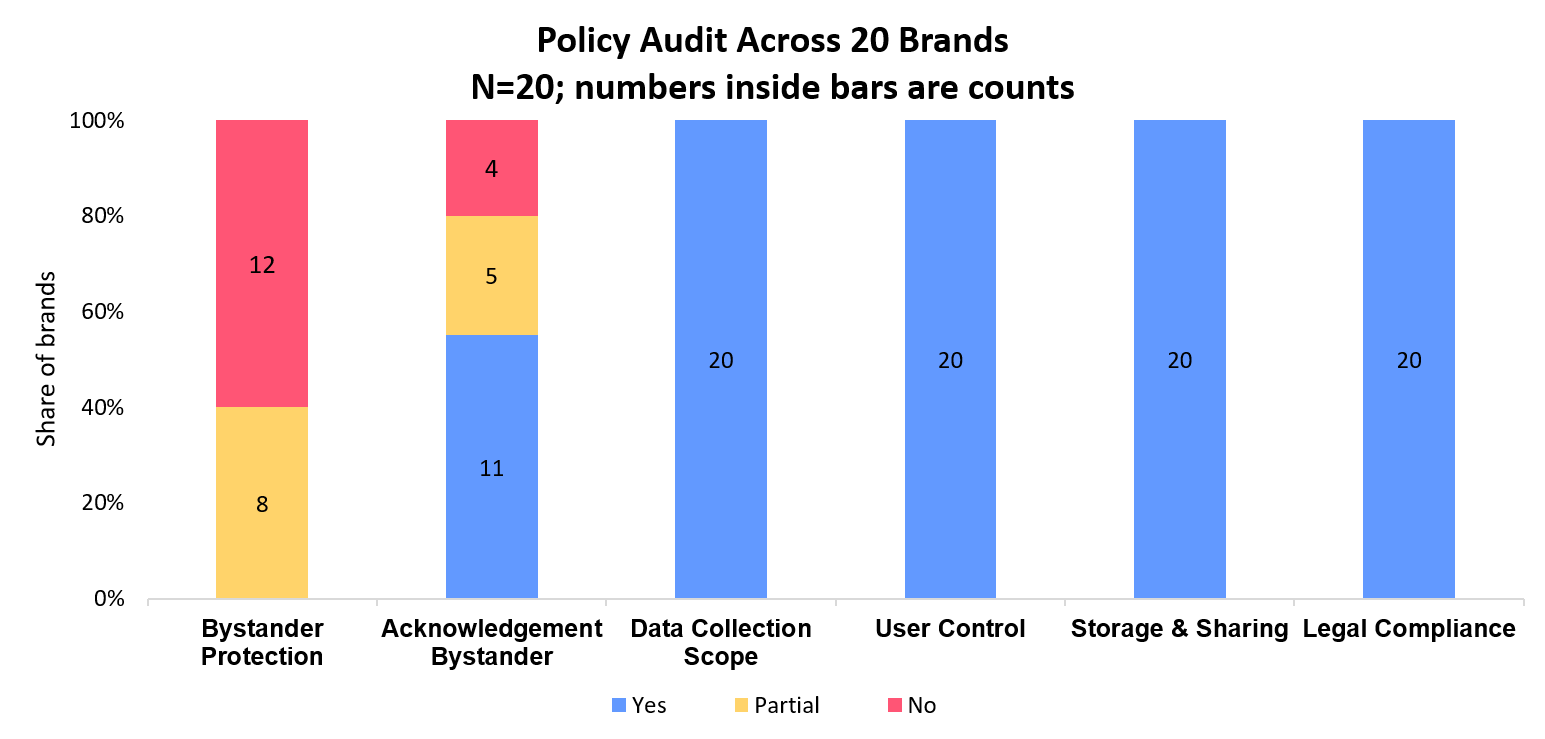}
  \caption{Smart home privacy policy analysis across 20 brands}
    \label{fig:stat2}
    \vspace{-3mm}
\end{figure}

The analysis revealed significant variability among brands. Most companies provided detailed disclosures on user-centric privacy matters, such as the range of data collected, storage practices, sharing policies, user control capabilities, and compliance with global regulations like GDPR and CCPA. Data encryption and other security mechanisms were frequently discussed.
However, policies addressing interdependent privacy -- especially concerning non-users and bystanders -- were often insufficient. As shown in Fig.~\ref{fig:stat2}, only a minority of brands mention that their products could capture personal data from non-users. Even fewer offered practical safeguards or notification mechanisms for bystanders to be informed, to object, or to seek redress. 

Consequently, while robust protections for direct users are common, the majority of policies fail to meaningfully recognize non-users affected by these devices. As smart home devices become increasingly widespread, more individuals risk being unintentionally recorded without awareness, practical remedies, or protection. This omission shifts accountability onto device owners without providing necessary tools or guidance. Smart home manufacturers often employ privacy policies as instruments of legal (declaring themselves processors as opposed to joint controllers) and ethical disclaimer, explicitly shifting part of the compliance and the whole ethical burden as it concerns bystanders onto device owners. This strategy allows vendors to avoid direct liability and ethical responsibility for controlling personal but non-user data. Rather than embedding technical safeguards or notification systems, many companies merely advise users to notify bystanders, obtain consent, or comply with applicable law -- without providing concrete support or enforcement mechanisms. This reflects a common trend: privacy policies are often used to avoid liability rather than to protect the privacy of bystanders in interdependent data settings. Examples of such disclaimers across different brands are shown in Table~\ref{tab:disclaimer_policies}.

\begin{table}[t]
\centering
\caption{Smart home vendors using privacy policies as disclaimers}
\scriptsize
\renewcommand\arraystretch{1.2}
\begin{tabularx}{\linewidth}{@{}p{2.4cm}X@{}}
\toprule
\textbf{Brand} & \textbf{Disclaimer-style Policy Clause} \\
\midrule

Google Nest & ``You (not Nest) are responsible for ensuring that you comply with any applicable laws... you may need to display a notice that alerts visitors to your home and/or obtain their explicit consent depending on the purpose and the means of your processing activities.'' \\ \midrule

Arlo & ``If you choose to enable the Person Recognition feature on your device, we will collect face images and face prints. Depending on where your device is located, this feature may require that you provide notice to and obtain consent from people who may be recognized by your camera.'' \\ \midrule

TP-Link Kasa & ``Depending on your country/region, you may be required to inform your guest, visitor, client, etc. that you are using a camera to record audio and video.'' \\ \midrule

EZVIZ & ``If the personal information you submitted to us contains other individuals’ information, you warrant that you have obtained the approval of such individuals.'' \\ \midrule

SimpliSafe & ``You (and not SimpliSafe) are solely responsible for ensuring that the Recording complies with all applicable laws. For example, you may be required to display a notice at the Premises that alerts visitors... that you are using a security camera...'' \\ \midrule

YI Technology & ``Depending on your region, you may be required to inform your guest, visitor, client, or other non-YI users that you are using a security camera... You may need to consider their privacy rights before sharing...'' \\ \midrule

Netatmo & ``You must inform your visitors about the functioning and operation of the Security Cameras.'' \\ \midrule

Bosch & ``Where necessary, inform other people... about the use of your product in an appropriate manner... Obtain any consent which may be required.'' \\ \midrule

Samsung SmartThings & ``Privacy and video surveillance laws... may apply... Please ensure that you comply with applicable law... Capturing, recording or sharing video or audio content that involves other people may affect their privacy rights.'' \\ \midrule

ADT & ``If you provide us any information relating to another person, you represent that you have the authority to do so and to permit us to use the information in accordance with this policy.'' \\

\bottomrule
\end{tabularx}
\label{tab:disclaimer_policies}
\vspace{-3mm}
\end{table}

\section{Privacy policies: critique and recommendations}
\label{sec:ppanalysis}
Here, we identify key shortcomings in the language, structure, and enforcement of current smart home privacy policies, specifically in how they address IDP. Then we offer concrete recommendations for both policy development and system design, aimed at strengthening privacy protection for bystanders affected by smart home technology.

\subsection{Shortcomings}

Our detailed review of privacy policies across the smart home industry reveals several systemic shortcomings regarding IDP. These gaps not only undermine the protection of bystanders but may also place an unreasonable burden on device owners, all while leaving service providers legally insulated.

\noindent\textbf{Superficial and vague treatment of IDP. }
The first and most salient issue is that references to IDP in privacy policies are rare, brief, and difficult to identify. When mentioned, these statements are often embedded within broader disclaimers and lack explicit language addressing bystander rights or obligations. In most cases, users may not even recognize that the policy refers to the rights or data of others (see Table~\ref{tab:privacy_comparison_llncs_final}). This observation aligns with recent literature that notes the persistent invisibility of IDP in consumer-facing documents~\cite{frik2025cares}.

\noindent\textbf{Transferring accountability via disclaimers. }
A substantial proportion of privacy policies analyzed function primarily as disclaimers. Companies routinely assert that users are solely responsible for ensuring the lawful operation of their devices, especially concerning recording third parties. However, these disclaimers are presented without meaningful guidance or detail. For example, policies often state that the user “must comply with applicable laws” or “obtain consent from bystanders,” but offer no explanation on what this entails, nor do they address the legal ambiguity and practical difficulty in doing so. In effect, the policies shift both the ethical and legal responsibility to users, but do not empower them to meet these obligations.

\noindent\textbf{Opaque legal references without user guidance. }
Privacy policies frequently cite complex legal regimes such as the GDPR, CCPA, or the UK Data Protection Act, but do not explain which articles are relevant to users or how these statutes impact everyday device operation. For instance, policies mention compliance with ``all applicable laws'' or reference regulatory frameworks, but they do not enumerate the sections that pertain to bystander data collection or user obligations. As a result, users are unlikely to understand their legal responsibilities, while bystanders have no meaningful information about their rights. The language remains inaccessible and fails to translate legal theory into actionable advice.

\noindent\textbf{Lack of practical examples and actionable guidance. }
Most policies fail to provide real-world scenarios or actionable steps for managing IDP risks. There are few, if any, practical examples of when and how to obtain consent from bystanders, or how to handle disputes. This omission stands in contrast to regulatory guidance in other domains (e.g., GDPR’s ``privacy by design'' or UK ICO’s CCTV code of practice\footnote{https://ico.org.uk/for-organisations/uk-gdpr-guidance-and-resources/cctv-and-video-surveillance/}), which emphasizes practical implementation. Without concrete examples, owners/users lack the knowledge and tools to protect others’ privacy, and bystanders remain uninformed and unprotected.

\subsection{Recommendations for policy and design improvements}
Although researchers have proposed various mechanisms to mitigate bystander privacy issues -- e.g., detecting and obfuscating bystander faces in video recordings~\cite{saqib2025bystander} --, our findings indicate that these techniques might not be deployed in practice. This gap reflects limited legal mandates\footnote{improved regulation could mandate technical safeguards for bystander privacy if the vendor is a data processor or joint controller} and potentially high implementation costs. In line with this reality, we outline practical, actionable recommendations drawn from regulatory and ethics best practices and technical research in interdependent privacy.

\noindent\textbf{Clear statements on accountability. }
Privacy policies should provide explicit, understandable statements about the responsibilities of device owners regarding bystander privacy. Unlike conventional data sharing scenarios in which the company is the primary data controller and the user is the protected data subject, IDP contexts complicate these roles: the user becomes a \emph{de facto} ``amateur'' joint data controller~\cite{helberger2010little,SYMEONIDIS2018179,SalehzadehNiksiratUSENIXDAB-IDP} for others' data. Policies should make this role change explicit and transparent. For instance, companies could include prominent warnings and summary tables listing user obligations when recording or storing data about others. Note that such accountability is not equivalent to legal liability, given current regulations.

%\vspace{-1mm}
\noindent\textbf{Policies as privacy-enabling tools, not just disclaimers. }
Privacy policies could serve as resources that empower users to mitigate IDP risks, rather than shields against company liability~\cite{slepchuk2020informing}. For example, they could promote voluntary bystander-protection measures such as providing owners with physical ``Notice'' stickers to alert passers-by to surveillance, or integrating visual indicators (e.g., camera LEDs) to signal active recording. Such measures are not only feasible but already required by law in certain jurisdictions: for instance, under the UK Protection of Freedoms Act 2012 and the Data Protection Act 2018, visible signs must indicate the presence of CCTV cameras in public or semi-public areas\footnotemark[13]. Providing technical options such as privacy zones (areas not recorded) and offering user-friendly templates for consent forms would also lower barriers for responsible use.

\noindent\textbf{Contextualized legal guidance. }
Rather than citing entire regulations, privacy policies should identify and summarize the specific legal provisions that affect both users and bystanders. For instance, China’s new regulation on public video surveillance makes it unlawful to install cameras in public areas without explicit authorization or notification.\footnote{\url{https://www.gov.cn/zhengce/content/202502/content\_7003024.htm} (in Chinese)} A concise summary of such provisions, presented in plain language, would enable users to understand and comply.

\noindent\textbf{Offering practical use cases and best practices. }
Policies should offer concrete examples: ``If your camera faces a public walkway, you must post a notice and disable audio recording,'' or ``If you record visitors, you should inform them and provide contact details for inquiries.'' Drawing on guidelines\footnote{{https://ico.org.uk/for-organisations/uk-gdpr-guidance-and-resources/cctv-and-video-surveillance/}} from the UK Information Commissioner’s Office (ICO) and recent court rulings, companies can provide template scenarios and checklists, helping users translate policy into ethical practice.

\noindent\textbf{Proactive bystander protection features. }
Manufacturers should adopt technical features that safeguard bystanders, such as default notifications when recording is active, automatic masking of adjacent property, or scheduled ``privacy times'' when cameras are disabled. Already demonstrated for IDP scenarios like photo sharing~\cite{olteanu}, third-party apps~\cite{DBLP:journals/compsec/LiuB25}, and digital address books~\cite{SalehzadehNiksiratUSENIXDAB-IDP}, voluntary technical solutions may complement policy to provide layered protection~\cite{DBLP:conf/sp/BirrellRDLMHL24}. Naturally, these technical countermeasures should also be referenced in privacy-related documentation.

\section{Conclusion}
\label{sec: conclusion}
Our analysis reveals that some current smart home privacy policies (and supplemental documentation) acknowledge the issue of recording and sharing the personal data of bystanders. However, when bystanders are mentioned, it is typically buried in obscure disclaimers that attempt to shift legal and ethical responsibility entirely to the device owner. There are almost no proactive warnings, guidance, or practical mechanisms to help owners/users recognize and address their perceived obligations regarding bystanders. As a result, bystander privacy is mostly neglected, and affected individuals are likely to remain unaware and vulnerable. Existing policies fail to reflect the realities of smart environments, where technology decisions have far-reaching impacts beyond the device owner. Therefore, we provided clear recommendations for improved privacy policies.

\subsubsection*{Limitations and future work. }
This work is limited to English-language privacy policies from major brands, while regional practices and smaller manufacturers may present different patterns: the supervised usage of LLMs may enable us to do policy analysis at scale in future work. Legal interpretations and technology continue to evolve; thus, our findings may not capture the most recent regulatory changes. Furthermore, we did not carry out experimental research with respect to user awareness and attitudes in a real-world smart home environment; such a study is next on our list. Finally, we plan to develop a ready-to-use privacy policy template for privacy-conscious vendors.

\bibliographystyle{unsrt}
\bibliography{reference}
\end{document}